\documentclass[twocolumn,usenatbib]{mn2e}
\usepackage{amsmath}
\usepackage{epsfig}
\usepackage{graphicx}
\usepackage{capt-of}
\usepackage{amssymb}
\begin{document}

\title[Fractional Derivative and Self-gravitation Equation]{Fractional Derivative Approach to the Self-gravitation Equation}

\author[J. F. Pedraza, J. Ramos-Caro and G. A. Gonz\'alez]
{Juan F. Pedraza$^{1}$\thanks{E-mail: jfpa080@gmail.com}, Javier F.
Ramos-Caro$^{1}$\thanks{E-mail: javiramos1976@gmail.com} and
Guillermo A. Gonz\'alez$^{1,2}$\thanks{E-mail: guillego@uis.edu.co} \\
$^{1}$Escuela de F\'isica, Universidad Industrial de Santander, A. A. 678,
Bucaramanga, Colombia\\
$^{2}$Departamento de F\'isica Te\'orica, Universidad del Pa\'is Vasco, 48080
Bilbao, Spain}

\maketitle

\begin{abstract}
A new formalism is presented for finding equilibrium distribution functions for
axisymmetric systems. The formalism, obtainded by using the concept of
fractional derivatives, generalizes the methods of \cite{fricke}, \cite{KAL} and
\cite{jiang}, and has the advantage that can be applied to a wider variety of
models. We found that this approach can be applied both to tridimensional
systems and to flat systems, without the necessity of dealing with a
pseudo-volume mass density. As an application, we obtain the distribution
functions of the Binney's logarithmic model and of the Mestel disc.
\end{abstract}

\begin{keywords}
stellar dynamics -- galaxies: kinematics and dynamics.
\end{keywords}

\section{Introduction}

The construction of self-consistent models for stellar systems is of great
interest in astrophysics. As it was pointed out by some authors, the most
straightforward way to perform such models is to start with an assumed potential
defining the mass density $\rho$ (via Poisson's equation) and the families of
orbits that can lie within the system (via Newton's equations of motion). Since
$\rho$ is the integration of the distribution function (DF) over the velocity
variables in the phase space of the system, the problem of finding the DF is
that of solving an integral equation. For that reason, such procedure is called
as the ``from $\rho$ to $f$'' approach for finding a self-consistent DF
(\cite{BT}; \cite{Hunter}; \cite{jiang}) and the integral relation connecting
$f$ and $\rho$ is known as the self-gravitation equation.

By Jeans's theorem, an equilibrium DF is a function of the isolating integrals
of motion that are conserved in each orbit. It has been shown that, for certain
potential-density pairs (PDP), it is possible to find analytically such kind of
DFs. The simplest case of physical interest corresponds to spherically symmetric
PDP, described by an isotropic DF that only depends on the energy. Eddington
(1916) showed that it is possible to obtain such DFs by first expressing the
density as a function of the potential, and then solving an Abel integral
equation.

Another case of interest in astrophysics is the corresponding to axially
symmetric systems, for which a great variety of PDP has been constructed, e.g.
\cite{KUZ}; \cite{T1}; \cite{Miya}; \cite{bagin}; \cite{KAL}; \cite{Miyamoto};
\cite{Nagai}; Kutuzov \& Ossipkov (1980, 1986, 1988); Evans (1993, 1994);
Kutuzov (1995); Jiang (2000); Jiang \& Moss (2002); \cite{GR}; Ossipkov \& Jiang
(2007). Now then, the first approach to obtain axisymmetric DFs was introduced
by \cite{fricke} and, after that, different integral transformation techniques
were used to obtain two-integral DFs (e.g. Lynden-Bell (1962); Hunter (1975);
Kalnajs (1976); Dejonghe (1986); Hunter \& Qian (1993)) but there appear several
obstacles concerning to the validity of such transformations and the requirement
of the complex analyticity of a density-related integral kernel to complex
arguments.

Recently, \cite{jiang} presented a new method for the axially symmetric case,
where the equilibrium DF depends on the energy $E$ and the angular momentum
about the axis of symmetry $L_{z}$, i.e. the two classical integrals of motion.
They developed a formalism that essentially combines both the Eddington formulae
and the Fricke (1952) expansion in order to obtain the DF's even part, starting
from a density that can be expressed as a function of the radial coordinate and
the gravitational potential. Thus, for a given $\rho$, the corresponding even DF
can be obtained by solving an Abel integral equation. Once such even part is
determined, one can find the DF's odd part by introducing some reasonable
assumptions about the mean circular velocity or using the maximum entropy
principle.

In the present paper we show an extension of the formulae derived earlier by
\cite{fricke} and \cite{jiang}, by introducing the fractional derivative
concept. The approach developed here has several advantages over the methods
introduced before. In one hand, the mathematical difficulties involved in the
formalisms based on transformation techniques can be easily overcome. It is
worth to point out that the formalism introduced by \cite{jiang} demands the
definition a volumetric pseudo-density, in order to be applicable to flat
systems. On the other hand, our method can be applied directly to the case of
flat systems.

Assume that $\Phi$ and $E$ are, respectively, the gravitational potential and
the energy of a star in a stellar system. It is useful to define a relative
potential $\Psi=-\Phi+\Phi_0$ and a relative energy $\varepsilon=-E+\Phi_0$, in
such a way that the system has only stars with energy $\varepsilon> 0$
(\cite{BT}). For the case of an axially symmetric system, it is customary to use
cylindrical polar coordinates $(R,\varphi,z)$, where the velocity is denoted by
${\bf v}=(v_R,v_\varphi,v_z)$. Such system admits two isolating integrals: the
component of the angular momentum about the $z$-axis, $L_z=R v_\varphi$, and the
relative energy $\varepsilon$. Hence, the DF of a steady-state stellar system in
an axisymmetric potential can be expressed as a non-negative function of
$\varepsilon$ and $L_z$, that vanishes for $\varepsilon<0$,  denoted by
$f(\varepsilon,L_z)$ and related to the mass density as
\begin{equation}
\rho = \frac{4\pi}{R} \int_0^\Psi \int_0^{R \sqrt{2(\Psi - \varepsilon)}}
f_+(\varepsilon,L_z) dL_z d\varepsilon, \label{inta2}
\end{equation}
where $f_{+}(\varepsilon,L_z)$ is the even part of $f$ with respect to the
angular momentum $L_z$. On the other hand, for the case of flat systems, the
surface mass density $\Sigma$ is related to $f$ through
\begin{equation}
\Sigma = 4 \int_0^\Psi \int_0^{R \sqrt{2 (\Psi - \varepsilon)}}
\frac{f_+(\varepsilon,L_z) dL_z d\varepsilon}{\sqrt{ 2 R^{2}(\Psi - \varepsilon)
- L_{z}^{2}}}. \label{inta3}
\end{equation}
Now, by defining a pseudo-volume density $\hat{\rho}$, according to (Hunter and
Quian (1993))
\begin{equation}
\hat{\rho} = \sqrt{2} \int_{0}^{\Psi} \frac{\Sigma (R^{2},\Psi')
d\Psi'}{\sqrt{\Psi - \Psi'}},\label{seudorho}
\end{equation}
it is also possible to use  equation (\ref{inta2}) to deal with these flat
systems.

\section{Tridimensional systems\label{3dform}}

Most of the methods developed to solve (\ref{inta2}) require some kind of
dependence between the DF, the relative energy $\varepsilon$ and the angular
momentum $L_z$ (see \cite{fricke} and \cite{jiang}, as examples), which will
define the corresponding mass density $\rho(R,\Psi)$. Therefore, in order to
study the problem, we shall start by assuming some general types of DFs.

\subsection{DFs of the form $\sum\limits_n L_z^{2\alpha_n} h_n(\varepsilon)$
\label{form1}}

To generalize the Jiang \& Ossipkov method, we suppose that the DF can be
written as
\begin{equation}
f_{+}(\varepsilon,L_{z}) = \sum_{n} L_{z}^{2\alpha_n} h_{n}(\varepsilon), \label{assumption2}
\end{equation}
where $\alpha_n\in\mathbb{R}$ and the $2$ in the exponent of $L_z$
guarantees that $f_+$ is even. Now, performing the integral
(\ref{inta2}) with respect to $L_z$, we obtain
\begin{equation}
\rho = \sum \limits_{n} \frac{4 \pi 2^{\alpha_n + \frac{1}{2}}
R^{2\alpha_n}}{2\alpha_n + 1} \int_0^\Psi h_n(\varepsilon)
(\Psi-\varepsilon)^{\alpha_n + \frac{1}{2}} d\varepsilon,
\label{eqshi}
\end{equation}
for $\alpha_n > -1/2$, while it diverges for $\alpha_n \leq -1/2$. Therefore, we
assume that the the mass density is given by
\begin{equation}
\rho(R,\Psi) = \sum_{n} R^{2\alpha_n} \tilde{\rho}_{n} (\Psi),\quad \mathrm{for}
\quad \alpha_n > - \frac{1}{2}, \label{rhoform}
\end{equation}
A comparison  between (\ref{eqshi}) and (\ref{rhoform}), leads to
the relation
\begin{equation}
\tilde{\rho}_n (\Psi) = \frac{4 \pi 2^{\alpha_n + \frac{1}{2}}}{2 \alpha_n + 1}
\int_0^\Psi h_n(\varepsilon) (\Psi - \varepsilon)^{\alpha_n + \frac{1}{2}}
d\varepsilon. \label{relation1}
\end{equation}
At this point, we introduce the fractional derivative operator $D_x^j$, which
represents a $j$-order derivative, with respect to $x$, for any real value of
$j$ (see \cite{fract}). Assuming that
$(D_\Psi^j\tilde{\rho}_n(\Psi))_{\Psi=0}=0$ for all $j\in(0,\alpha_n+1/2)$, then
\begin{equation}
D_\Psi^{\alpha_n + \frac{1}{2}} \tilde{\rho}_{n} (\Psi) = \pi 2^{\alpha_n +
\frac{3}{2}} \Gamma(\alpha_n + \frac{1}{2}) \int_{0}^{\Psi} h_{n} (\varepsilon)
d\varepsilon. \label{refeq}
\end{equation}
This integral equation is simpler than the first one and can be inverted easily
if one takes the derivative once again with respect to $\Psi$,
\begin{equation}
h_{n}(\varepsilon) = \frac{\left. D_\Psi^{\alpha_n + \frac{3}{2}} \tilde{\rho}_{n} (\Psi)
\right|_{\Psi = \varepsilon}}{\pi 2^{\alpha_n + \frac{3}{2}} \Gamma(\alpha_n +
\frac{1}{2})},
\end{equation}
and the distribution function can be expressed as
\begin{equation}
f_{+}(\varepsilon,L_{z}) =  \sum_{n} \frac{L_{z}^{2\alpha_n} \left.
D_\Psi^{\alpha_n + \frac{3}{2}} \tilde{\rho}_{n}(\Psi)
\right|_{\Psi=\varepsilon}}{\pi 2^{\alpha_n + \frac{3}{2}} \Gamma(\alpha_n +
\frac{1}{2})}. \label{dfnew}
\end{equation}
When $\alpha_n\in\mathbb{N}$, by the definition of the Riemann-Liouville
operator, equation (\ref{dfnew}) reduces to the formulae obtained by
\cite{jiang}.

As a particular case, suppose that $\tilde{\rho}_{n}(\Psi)$ can be written in
the form
\begin{equation}
\tilde{\rho}_{n}(\Psi) = \sum_k A_{nk} \Psi^{\beta_k}. \label{tildero}
\end{equation}
So, applying the fractional derivative operator to (\ref{tildero})
\begin{equation}
D_\Psi^{\alpha_n + \frac{3}{2}} \tilde{\rho}_{n}(\Psi) = \sum_k \frac{A_{nk}
\Gamma(\beta_k + 1) \Psi^{\beta_k -
\alpha_n - \frac{3}{2}}}{\Gamma(\beta_k - \alpha_n - \frac{1}{2})},
\end{equation}
for $\beta_k > \alpha_n + 1/2$, and the corresponding DF is
\begin{equation}
f_{+} = \sum_{n,k} \frac{A_{nk} \Gamma(\beta_k + 1)
L_{z}^{2\alpha_n} \varepsilon^{\beta_k - \alpha_n - \frac{3}{2}}}{\pi
2^{\alpha_n + \frac{3}{2}} \Gamma(\alpha_n + \frac{1}{2}) \Gamma(\beta_k -
\alpha_n - \frac{1}{2})}.
\end{equation}
This result is totally equivalent to the Fricke solution, for real
values of $\alpha_n$, and therefore can be considered as a generalization.

\subsection{DFs of the form $\sum\limits_n L_z^{2\alpha_n} g_n(Q)$
\label{dfgenq}}

It is possible to derive a more general expression for the DF if we assume that
it depends on $\varepsilon$ through $Q=\varepsilon-L_z^2/(2R_a^2)$, where $R_a$
is a scaling radius. Suppose that the system has only stars with $Q>0$, so
$f(Q,L_z)=0$ for $Q\leq0$. Here, $Q\rightarrow\varepsilon$ as
$R_a\rightarrow\infty$. The fundamental equation can be written, in terms of
$Q$, as
\begin{equation}
\rho = \frac{4 \pi}{R} \int_0^{\Psi} \int_0^{R \sqrt{2(\Psi - Q)/(1 +
\frac{R^2}{R_a^2})}} f_{+}(Q,L_z) dL_z dQ, \label{eqintq}
\end{equation}
where $f_{+}(Q,L_z)$ is the even part of $f(Q,L_z)$. So, following a similar
procedure than in section \ref{form1}, one can find that a DF of the form
\begin{equation}
f_{+}(Q,L_{z}) = \sum_{n} \frac{L_{z}^{2\alpha_n} \left. D_\Psi^{\alpha_n +
\frac{3}{2}} \tilde{\rho}_{n}(\Psi) \right|_{\Psi=Q}}{\pi 2^{\alpha_n +
\frac{3}{2}} \Gamma(\alpha_n + \frac{1}{2})},
\end{equation}
corresponds to a mass density of the form
\begin{equation}
\rho(R,\Psi) = \sum_{n} \frac{R^{2\alpha_n} \tilde{\rho}_{n}(\Psi)}{(1 + \frac{R^2}{R_a^2})^{\alpha_n + \frac{1}{2}}} ,\label{rhoformq}
\end{equation}
for $\alpha_n>-1/2$, where $\alpha_n\in\mathbb{R}$. Now, if we sum over all
posible values of $R_a$ we obtain the general expression
\begin{equation}
f_{+}(Q,L_{z}) = \sum_{a,n} \frac{L_{z}^{2\alpha_n} \left. D_\Psi^{\alpha_n +
\frac{3}{2}} \tilde{\rho}_{n}(\Psi) \right|_{\Psi = Q}}{\pi2^{\alpha_n +
\frac{3}{2}} \Gamma(\alpha_n + \frac{1}{2})},
\end{equation}
corresponding to a density of the form
\begin{equation}
\rho(R,\Psi) = \sum_{a,n} \frac{R^{2\alpha_n} \tilde{\rho}_{n}(\Psi)}{(1 +
\frac{R^2}{R_a^2})^{\alpha_n + \frac{1}{2}}},
\end{equation}
with $R_a>0$ and $\alpha_n>-1/2$.

\subsection{Models with divergent gravitational potential}

In a system in which the gravitational potential has no upper bound,
it is not possible to define correctly the relative potential $\Psi$
and the relative energy $\varepsilon$, because the scape energy of
the system is $\infty$. For this reason, we shall write the
fundamental equation in terms of $E$ and $\Phi$,
\begin{equation}\label{eqintnl}
\rho(R,\Phi) = \frac{4\pi}{R} \int_\Phi^{\infty} \int_0^{R\sqrt{2(E - \Phi)}}
f_{+}(E,L_z) dL_z dE,
\end{equation}
and we will suppose that the DF can be written as
\begin{equation}
f_{+}(E,L_{z}) = \sum_{n} L_{z}^{2\alpha_n} h_{n}(E), \label{assum}
\end{equation}
for $\alpha_n > -1/2$, and that the density is given by
\begin{equation}
\rho(R,\Phi) = \sum_{n} R^{2\alpha_n} \tilde{\rho}_{n}(\Phi), \label{rhoform2}
\end{equation}
for $\alpha_n > -1/2$. Thus then, by integrating with respect to $L_z$, follows
that 
\begin{equation}
\tilde{\rho}_n(\Phi) = \frac{4\pi 2^{\alpha_n + \frac{1}{2}}}{2\alpha_n + 1}
\int_\Phi^\infty h_n(E) (E - \Phi)^{\alpha_n + \frac{1}{2}} dE. \label{relatinf}
\end{equation}
Now, if we assume that
\begin{equation}
\lim_{\Phi \rightarrow \infty} D_\Phi^j \tilde{\rho}_n(\Phi)=0\label{asum}
\end{equation}
for all $j\in(0,\alpha_n + 1/2)$, then
\begin{equation}
D_\Phi^{\alpha_n + \frac{1}{2}} \tilde{\rho}_{n} = i \pi(-2)^{\alpha_n} 2^{\frac{3}{2}} \Gamma(\alpha_n + \frac{1}{2})
\int_{\Phi}^{\infty} h_{n}(E) dE, \label{eqq223}
\end{equation}
and so we obtain
\begin{equation}
h_{n}(E) = \frac{\left. D_\Phi^{\alpha_n + \frac{3}{2}} \tilde{\rho}_{n}(\Phi)
\right|_{\Phi = E}}{\pi (-2)^{\alpha_n + \frac{3}{2}} \Gamma(\alpha_n +
\frac{1}{2})}.
\end{equation}
Therefore, the distribution function is
\begin{equation}
f_{+}(E,L_{z}) = \sum_{n} \frac{L_{z}^{2\alpha_n} \left. D_\Phi^{\alpha_n +
\frac{3}{2}} \tilde{\rho}_{n}(\Phi) \right|_{\Phi = E}} {\pi (-2)^{\alpha_n +
\frac{3}{2}} \Gamma(\alpha_n + \frac{1}{2})}, \label{dfnewinf}
\end{equation}
for $\alpha_n > - 1/2$.

On the other hand, we can assume that the density has the more general form
\begin{equation}
\rho(R,\Phi) = \sum_{n} \frac{R^{2\alpha_n} \tilde{\rho}_{n}(\Phi)}{(1 +
\frac{R^2}{R_a^2})^{\alpha_n + \frac{1}{2}}}, \label{rhoform2q}
\end{equation}
for $\alpha_n>-1/2$. So, the corresponding DF will be
\begin{equation}
f_{+}(Q,L_{z}) = \sum_{n} \frac{L_{z}^{2\alpha_n} \left. D_\Phi^{\alpha_n +
\frac{3}{2}} \tilde{\rho}_{n}(\Phi) \right|_{\Phi = Q}}{\pi (-2)^{\alpha_n +
\frac{3}{2}} \Gamma(\alpha_n + \frac{1}{2})}, \label{dfnewinfq}
\end{equation}
for $\alpha_n>-1/2$ and $Q$ defined as $Q=E+L_z^2/(2R_a^2)$. Now, if we sum
over all the posible values of $R_a$, we can obtain the generalization
\begin{equation}
f_{+}(Q,L_{z}) = \sum_{a,n} \frac{L_{z}^{2\alpha_n} \left. D_\Phi^{\alpha_n +
\frac{3}{2}} \tilde{\rho}_{n}(\Phi)\right|_{\Phi = Q}}{\pi (-2)^{\alpha_n +
\frac{3}{2}} \Gamma(\alpha_n + \frac{1}{2})},
\end{equation}
corresponding to a density of the form
\begin{equation}
\rho(R,\Phi) = \sum_{a,n} \frac{R^{2\alpha_n} \tilde{\rho}_{n}(\Phi)}{(1 +
\frac{R^2}{R_a^2})^{\alpha_n + \frac{1}{2}}},
\end{equation}
with $R_a > 0$ and $\alpha_n > -1/2$.

\section{Flat systems\label{2dform}}

The formalism sketched above can also be used directly in the case of flat
systems. Note that in the method introduced by \cite{jiang} it was not possible,
since the fundamental equation could not be solved using the Abel integral
equation. Now, we will proceed similarly to the tridimensional case, finding the
DFs for different kinds of densities. Then, we will also study the case of
models with divergent gravitational potential.

\subsection{DFs of the form $\sum\limits_n L_z^{2\alpha_n} h_n(\varepsilon)$ \label{dfplane}}

As in the tridimensional case, first we suppose that
\begin{equation}
f_{+}(\varepsilon,L_{z}) = \sum_{n} L_{z}^{2\alpha_n} h_{n}(\varepsilon).
\label{assumption}
\end{equation}
So, by integrating (\ref{inta3}) with respect to $L_z$, we obtain
\begin{equation}
\Sigma = \sum_{n} \frac{\sqrt{\pi} R^{2\alpha_n} \Gamma(\alpha_n + \frac{1}{2})}
{2^{-(\alpha_n + 1)} \Gamma(\alpha_n + 1)} \int_{0}^{\Psi}(\Psi -
\varepsilon)^{\alpha_n} h_{n} (\varepsilon) d\varepsilon, \label{newsigma2}
\end{equation}
for $\alpha_n>-1/2$. Therefore, if we assume that
\begin{equation}
\Sigma(R,\Psi)=\sum_{n}R^{2\alpha_n}\sigma_{n}(\Psi), \label{newsigma3}
\end{equation}
$\alpha_n > -1/2$, it follows that
\begin{equation}
\sigma_{n}(\Psi) = \frac{\sqrt{\pi} \Gamma(\alpha_n + \frac{1}{2})}
{2^{-(\alpha_n + 1)} \Gamma(\alpha_n + 1)} \int_{0}^{\Psi}(\Psi -
\varepsilon)^{\alpha_n} h_{n}(\varepsilon) d\varepsilon. \label{newsigma4}
\end{equation}
Now, if $(D_\Psi^j\sigma_n(\Psi))_{\Psi = 0} = 0$ for all $j \in (0,\alpha_n)$,
then equation (\ref{newsigma4}) leads to
\begin{equation}
D_\Psi^{\alpha_n} \sigma_{n}(\Psi) = \sqrt{\pi} 2^{\alpha_n+1} \Gamma(\alpha_n +
\frac{1}{2}) \int_{0}^{\Psi} h_{n}(\varepsilon) d\varepsilon. \label{newsigma5}
\end{equation}
Consequently,
\begin{equation}
h_{n}(\varepsilon) = \frac{\left. D_\Psi^{\alpha_n + 1} \sigma_{n}(\Psi)
\right|_{\Psi = \varepsilon}}{\sqrt{\pi} 2^{\alpha_n + 1}\Gamma(\alpha_n +
\frac{1}{2})}, \label{newsigma6}
\end{equation}
and the DF is
\begin{equation}
f_{+}(\varepsilon,L_{z}) = \sum_{n} \frac{L_{z}^{2\alpha_n} \left.
D_\Psi^{\alpha_n + 1} \sigma_{n}(\Psi) \right|_{\Psi = \varepsilon}}{\sqrt{\pi}
2^{\alpha_n + 1} \Gamma(\alpha_n + \frac{1}{2})}. \label{dftotal}
\end{equation}
This equation, for $\alpha_n=0$, corresponds to the method developed in
\cite{kal}, working in an adequate rotating frame.

As a particular case, suppose that
\begin{equation}
\sigma_{n}(\Psi) = \sum_k A_{nk} \Psi^{\beta_k}.
\end{equation}
Then, by taking the fractional derivative we obtain
\begin{equation}
D_\Psi^{\alpha_n + 1} \sigma_{n}(\Psi) = \sum_k \frac{A_{nk}
\Gamma(\beta_k + 1) \Psi^{\beta_k - \alpha_n - 1}}{\Gamma(\beta_k - \alpha_n)},
\end{equation}
and the DF is
\begin{equation}
f_{+}(\varepsilon,L_{z}) = \sum_{n,k} \frac{A_{nk} \Gamma(\beta_k + 1) L_{z}^{2\alpha_n} \varepsilon^{\beta_k - \alpha_n - 1}}{\sqrt{\pi} 2^{\alpha_n + 1} \Gamma(\alpha_n + \frac{1}{2}) \Gamma(\beta_k - \alpha_n)}.
\end{equation}
This relation can be interpreted as the analogous case of the Fricke expansion,
when we are dealing with flat systems. It can be verified performing the
pseudo-volume density (\ref{seudorho}) of $R^{2\alpha_n}\Psi^{\beta_k}$ and
taking the Frike component corresponding to the tridimensional case.

\subsection{Other DFs for flat systems\label{dfgenq2}}

We can generalize the result (\ref{dftotal}) if we express the DF in terms of $Q
= \varepsilon-L_z^2/(2R_a^2)$. In this way, if the density has the form
\begin{equation}
\Sigma(R,\Psi) = \sum_{a,n} \frac{R^{2\alpha_n} \sigma_{n}(\Psi)}{(1 +
\frac{R^2}{R_a^2})^{\alpha_n}},
\end{equation}
the corresponding DF is
\begin{equation}
f_{+}(Q,L_{z}) = \sum_{a,n} \frac{L_{z}^{2\alpha_n} \left. D_\Psi^{\alpha_n + 1}
\sigma_{n}(\Psi) \right|_{\Psi = Q}}{\sqrt{\pi} 2^{\alpha_n + 1}\Gamma(\alpha_n
+ \frac{1}{2})}, \label{dftotalqp2}
\end{equation}
for $R_a > 0$ and $\alpha_n > -1/2$.

Furthermore, if we consider models with gravitational potential having no upper
bound, we can deduce that for a density
\begin{equation}
\Sigma(R,\Phi) = \sum_{a,n} \frac{R^{2\alpha_n} \sigma_{n}(\Phi)}{(1 +
\frac{R^2}{R_a^2})^{\alpha_n + \frac{1}{2}}},
\end{equation}
and assuming that $\lim_{_{\Phi \rightarrow \infty}} D_\Phi^j \sigma_n(\Phi) =0$ 
for all $j\in(0,\alpha_n)$, then
\begin{equation}
f_{+}(Q,L_{z}) = \sum_{a,n} \frac{L_z^{2\alpha_n} \left. D_\Phi^{\alpha_n + 1}
\sigma_{n}(\Phi) \right|_{\Phi = Q}}{\sqrt{\pi} (-2)^{(\alpha_n + 1)}
\Gamma(\alpha_n + \frac{1}{2})}, \label{nouppf}
\end{equation}
for $R_a > 0$, $\alpha_n > -1/2$ and $Q = E + L_z^2/(2R_a^2)$.

\section{SOME APPLICATIONS\label{applic}}

In this section we will use the formulae introduced above to the Binney's
logarithmic model and the Mestel disc, and we will see that their corresponding
DFs match exactly with those that were found through the application of other
methods. Binney's logarithmic model has a gravitational potential of the form
\begin{equation}
\Phi(R,z) = \frac{1}{2} v_0^2 \ln \left(1 + R^2 + \frac{z^2}{q^2} \right),
\label{potbinney}
\end{equation}
whereas its mass density is
\begin{equation}
\rho(R,z) = \frac{v_0^2 (1 + 2 q^2 + R^2 + (2 - q^{-2}) z^2)}{4 \pi G q^2 (1 +
R^2 + z^2 q^{-2})^2}, \label{denbinney}
\end{equation}
that can be written as
\begin{equation}\label{denbinney2}
\rho = \frac{v_0^2 \{ [(1 - q^2) R^2 + 1] + (q^2 - \frac{1}{2}) e^{2\Phi/v_0^2}
\}}{2 \pi G q^2 e^{4\Phi/v_0^2}}.
\end{equation}
So, as $D_x^\alpha e^{ax}=a^\alpha e^{ax}$ for any $\alpha\in\mathbb{R}$, we
obtain
\begin{equation}
f_{+}(E,L_z)=A L_z^2e^{-4E/v_0^2}+B e^{-4E/v_0^2}+C
e^{-2E/v_0^2},\label{dfbinney}
\end{equation}
where
$$
A = \frac{2^{\frac{5}{2}}(1-q^2)}{\pi^{\frac{5}{2}} G q^2 v_0^3}, \quad
B = \frac{\sqrt{2}}{\pi^{\frac{5}{2}} G q^2 v_0}, \quad
C = \frac{2q^2-1}{4\pi^{5/2}Gq^2v_0},
$$
the same DF founded in \cite{jiang}, using the Abel's integral equation, and in
\cite{eva93b} using Lynden-Bell's method.

Another case of interest is the Mestel disc, characterized by a gravitational
potential of the form
\begin{equation}
\Phi(R) = v_c^2 \ln \left(\frac{R}{R_0}\right), \label{potmestel}
\end{equation}
and a surface mass density given by
\begin{equation}
\Sigma(R) = \Sigma_0 \frac{R_0}{R}, \label{denmestel}
\end{equation}
where $\Sigma_0=v_c^2/(2 \pi G R_0)$. Now, we can write (\ref{denmestel}) as
\begin{equation}
\Sigma(R)=R^{2m}\frac{\Sigma_0}{R_0^{2m}}e^{-(2m+1)\Phi/v_c^2},
\end{equation}
for any $m\in\mathbb{R}$. So, equation (\ref{nouppf}) for
$R_a\rightarrow\infty$ leads to
\begin{equation}
f_{+}(E,L_z)=F L_z^{2m}e^{-E/\sigma^2},
\end{equation}
where $F$ and $\sigma$ are the constants given by
\begin{equation}
\sigma^2=\frac{v_c^2}{2m+1}, \quad
F=\frac{\Sigma_02^{-(m+1)}\pi^{-1/2}}{\Gamma(m+1/2)R_0^{2m}\sigma^{2m+2}}.
\end{equation}
This solution was obtained as well by \cite{eva93b}, and was proposed earlier 
by \cite{T3}.

\section{DISCUSSION}

In contrast with the methods based on integral transformation techniques, the
formalism developed here does not require that the mass density has an analytic
continuation to complex arguments. Indeed, this is the principal disadvantage
involved in such methods. Moreover, our fractional derivative approach can be
regarded as a general method that contains, as particular cases, the results
obtained by \cite{fricke}, \cite{kal} and \cite{jiang}. The method developed
here can be applied to a wider variety of axisymmetric models, due to the
generic form of the density  as a function. Another advantage of this formalism
is that it can be applied directly both to tridimensional systems and to flat
systems, without the implementation of a pseudo-volume density. Therefore,
taking into account all the above statements, the present formalism represents a
powerful tool on the construction of self-consistent stellar models.

\section{Acknowledgments}

J. R-C. wants to thank the financial support from {\it Vicerrector\'ia
Acad\'emica}, Universidad Industrial de Santander.

\end{document}